\documentclass[a4paper,10pt]{article}
\usepackage[a4paper, margin=2cm]{geometry}

\usepackage[utf8]{inputenc}
\usepackage[T1]{fontenc}

\usepackage{fourier}

\usepackage[square,numbers]{natbib}
\usepackage{amsmath}
\allowdisplaybreaks
\usepackage{amsfonts}
\usepackage{bm}
\usepackage{mathrsfs}
\usepackage{booktabs}
\usepackage{graphicx}
\usepackage{url}
\usepackage{xcolor}

\usepackage{amsthm}
\theoremstyle{remark}
\newtheorem*{remark}{Remark}

\usepackage[english]{babel}
\usepackage[hidelinks]{hyperref}

\usepackage{doi}

\usepackage{ulem} 

\title{Models of fractional viscous stresses for incompressible materials}
\author{Harold Berjamin \textsuperscript{a}, Michel Destrade \textsuperscript{a,b} \\ ~ \\
\emph{\footnotesize\textsuperscript{a}School of Mathematical and Statistical Sciences, University of Galway, University Road, Galway, Republic of Ireland}\\
\emph{\footnotesize\textsuperscript{b}Key Laboratory of Soft Machines and Smart Devices of Zhejiang Province and Department of Engineering Mechanics,}\\ \emph{\footnotesize Zhejiang University, Hangzhou 310027, People's Republic of China}}

\date{}

\begin{document}

	\maketitle
	
	\begin{abstract}
		\noindent
		We present and review several models of fractional viscous stresses from the literature, which generalise classical viscosity theories to fractional orders by replacing total strain derivatives in time with fractional time derivatives. We also briefly introduce  Prony-type approximations of these theories.
		Here we investigate the issues of material frame-indifference and thermodynamic consistency for these models and find that on these bases, some are physically unacceptable. Next we study elementary shearing and tensile motions, observing that some models are more convenient to use than others for the analysis of creep and relaxation.
		Finally, we compute the incremental stresses due to small-amplitude wave propagation in a deformed material, with a view to establish acousto-elastic formulas for prospective experimental calibrations.
		~ \\
		
	\end{abstract}

\section{Introduction}\label{sec:Intro}

\begin{figure}
	\centering
	\includegraphics[width=0.8\textwidth]{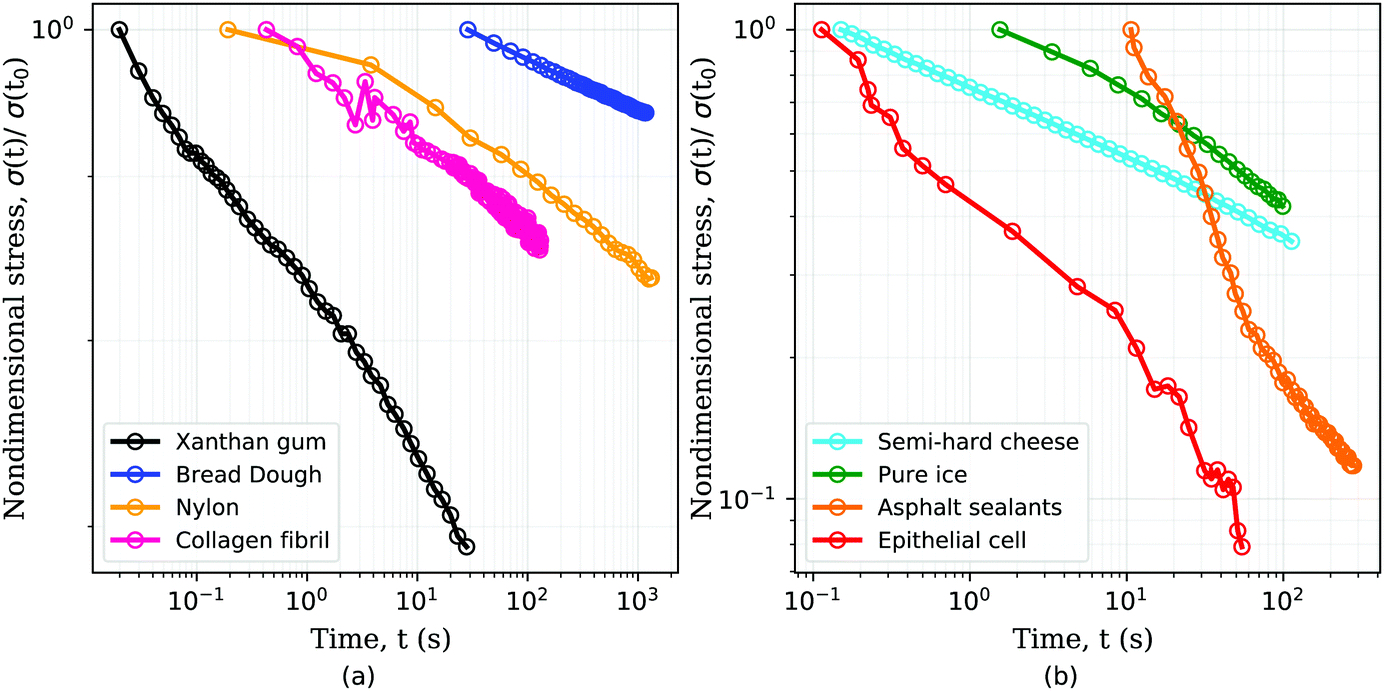}
	\caption{Stress relaxation data sets show that the response of many materials follows a power-law behaviour. Reproduced from Bonfanti et al.~\cite{bonfanti20} under the \href{https://creativecommons.org/licenses/by/3.0/}{Creative Commons Attribution 3.0 Unported Licence}. \label{fig-Bonfanti}}
\end{figure}

Differential operators based on the Riemann-Liouville integral are commonly used to generalise differentiation of integer order to fractional orders, thus laying the foundations of \emph{fractional calculus} \cite{matignon09}. This branch has found various applications over time \cite{podlubny99,tarasov19}, most notably in electronics and in material rheology. In the latter case \cite{mainardi10,atanackovic14}, fractional calculus provides accurate predictions of the time-dependent mechanical behaviour with a limited number of parameters.
Figure~\ref{fig-Bonfanti} lists a few materials whose stress relaxation response can be described by such theories, e.g., xanthan gum, bread dough, and nylon. One property of these materials is the power-law time evolution of the stress in response to a sudden deformation, see the review by Bonfanti et al.~\cite{bonfanti20} for details. A possible explanation of the micromechanical origins of fractional behaviour is provided by Brenner \cite{brenner19}.

Classical rheological models are described by an assembly of one-dimensional elements such as springs and dashpots. Within this framework, fractional viscous elements are represented by \emph{springpots} which are intermediates between springs (accounting for elasticity) and dashpots (accounting for viscous dissipation), see Mainardi \cite{mainardi10} for an overview of the linear theory valid at small strains. The simplest such model is based on a single springpot element, which can be found under the name of the `Kjartansson constant-$Q$ model' in geophysics \cite{carcione15}.

\begin{figure}
	\centering
	\includegraphics{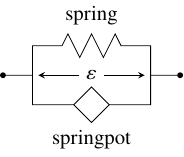}
	\caption{ Schematic representation of the fractional Kelvin-Voigt model. \label{fig-Diagram} }
\end{figure}

The mechanical analogue of Newtonian or Kelvin-Voigt viscoelastic materials is a dashpot connected to a spring in parallel. In this case, elastic stresses are proportional to the scalar strain $\varepsilon$, whereas viscous stresses are proportional to the velocity gradient or strain rate $\dot\varepsilon$ (in what follows, $\dot{\varepsilon}$ denotes the material time derivative of $\varepsilon$). The strain rate can be generalised to fractional differential orders $\alpha \in \left]0,1\right[$ based on the Caputo time-derivative $\mathfrak{D}^\alpha$ defined by
\begin{equation}
	\mathfrak{D}^\alpha \varepsilon(t) = \frac{1}{\Gamma(1-\alpha)} \int_{0}^t \frac{\dot{\varepsilon}(s)}{(t-s)^\alpha} \, \text d s ,
	\label{Frac}
\end{equation}
where $\varepsilon$ is causal and $\Gamma(z) = \int_0^\infty t^{z-1}\text{e}^{-t}\text{d}t$ defines Euler's Gamma function. This step amounts to replacing the dashpot with a springpot in the diagram, see Figure~\ref{fig-Diagram}.

With the above definition of the fractional derivative, we have the limits $\mathfrak{D}^0 \varepsilon = \varepsilon$ and, by differentiation, $\mathfrak{D}^1 \varepsilon = \dot \varepsilon$. Thus, $\alpha = 0$ corresponds to an elastic model, whereas $\alpha=1$ recovers the classical Kelvin-Voigt theory. Generalisation of \eqref{Frac} to arbitrary orders $\alpha\geq 0$ can be carried out \cite{freed07}, as well as generalisation to non-causal functions (e.g., periodic ones \cite{vigue19}). In this latter case, the Caputo derivative may be replaced by the Weyl derivative, which amounts to evaluating the integral \eqref{Frac} from $-\infty$ to $t$.

Fractional calculus has also been used to model the time-dependent mechanical response of highly-deformable soft materials for which the small strain assumption is no longer valid. Examples of such materials include amorphous polymers, organic glasses \cite{palade99}, rubber, elastomers \cite{drozdov97,haupt02}, tendons \cite{bologna20}, liver tissue \cite{capilnasiu20}, and other similar materials. While the generalisation of the spring to three-dimensional finite strain is rather straightforward (see the textbook by Holzapfel~\cite{holzapfel00} for a description of hyperelasticity), the crux of the matter is a proper definition of the fractional viscous stress.
In the present study, we consider the finite motion of \emph{incompressible} viscous materials. More specifically, we  describe nonlinear three-dimensional generalisations of the fractional Kelvin-Voigt rheology depicted in Figure~\ref{fig-Diagram}.

To describe their motion, we introduce $\bm{F}(t)$, the deformation gradient tensor at time $t$, which is defined as the gradient $\partial \bm{x}/\partial \bm{X}$ of the current position $\bm{x} = \bm{x}(t)$ of a particle with respect to its position $\bm{X} = \bm{x}(0)$ in the reference configuration. Incompressible materials do not allow for volume change. Hence, isochoricity is enforced at all times ($\det \bm{F} \equiv 1$), and the mass density $\rho$ is constant.

In a similar fashion to the linear scalar case, we assume that the second Piola-Kirchhoff stress tensor $\bm S$ and the Cauchy stress tensor $\bm{\sigma} = \bm{F}\bm{S}\bm{F}^\text{T}$ can be split additively into elastic and viscous parts, i.e.,
\begin{equation}
	\bm{S} = -p\bm{C}^{-1} +\bm{S}^\text{e}+\bm{S}^\text{v}, \qquad
	\bm{\sigma} = -p\bm{I} +\bm{\sigma}^\text{e}+\bm{\sigma}^\text{v} ,
	\label{Constitutive}
\end{equation}
where $\bm{C} = \bm{F}^\text{T}\bm{F}$ is the right Cauchy-Green deformation tensor, $\bm I$ is the identity tensor, and the scalar $p$ is a Lagrange multiplier accounting for the incompressibility constraint. The stresses with exponents \textsuperscript{e} and \textsuperscript{v} are elastic and viscous contributions, respectively. Elastic contributions vanish in the fluid limit, which can therefore be viewed as a special case of the present theories.

Up to a suitable redefinition of the pressure coefficient $p$, we note that the partial stress tensors $\bm{S}^{\bullet}$, $\bm{\sigma}^{\bullet}$ in Eq.~\eqref{Constitutive} can be replaced with their deviatoric counterparts $\bm{S}^{\bullet}_\text{D}$, $\bm{\sigma}^{\bullet}_\text{d}$. Here, the deviators are denoted by the subscripts `D' and `d', such that
\begin{equation}
	(\bullet)_\text{D} = (\bullet) - \tfrac13 \text{tr}[(\bullet) \bm{C}] \bm{C}^{-1} ,
	\qquad
	(\bullet)_\text{d} = (\bullet) - \tfrac13 \text{tr}(\bullet) \bm{I} ,
	\label{Deviator}
\end{equation}
and $\text{tr}$ is the trace. With these definitions, we note that the tensor $(\bullet)_\text{d}$ is trace-free.

In classical Newtonian viscosity theories, the viscous Cauchy stress is typically expressed as $\bm{\sigma}^\text{v} = 2\eta \bm{D}$ where $\eta > 0$ is the dynamic viscosity. The tensor $\bm{D} = \frac12 (\bm{L} + \bm{L}^\text{T})$ with $\bm{L} = \dot{\bm F}\bm{F}^{-1}$ is the Eulerian strain rate tensor, which is trace-free by virtue of incompressibility{\,---\,}in other words, $\bm{\sigma}^\text{v} = \bm{\sigma}^\text{v}_\text{d}$. In terms of the second Piola-Kirchhoff stress, Newtonian viscosity is then expressed by the relationship
\begin{equation}
\bm{S}^\text{v} = 2\eta \dot{\bm \Pi} = 2\eta \bm{C}^{-1} \dot{\bm E} \bm{C}^{-1} ,
\label{Newtonian}
\end{equation}
where
\begin{equation}
	\bm{\Pi} = \tfrac{1}{2} (\bm{I}-\bm{C}^{-1}), \qquad \bm{E} = \tfrac{1}{2}( \bm{C} - \bm{I}) ,
	\label{Strain}
\end{equation}
denote a Piola-type deformation tensor and the Green-Lagrange strain tensor $\bm{E} = \bm{C}\bm{\Pi}$, respectively. The equality $\dot{\bm \Pi} = \bm{C}^{-1} \dot{\bm E} \bm{C}^{-1}$ used in \eqref{Newtonian} follows from differential rules.

As an alternative to Newtonian viscosity \eqref{Newtonian}, one might define the viscous stress as
\begin{equation}
\bm{S}^\text{v} = 2\eta \dot{\bm E} ,
\label{GL} 
\end{equation}
which is proportional to the rate of Green-Lagrange strain.
In Physical Acoustics, this viscous term is sometimes preferred to the Newtonian viscous term, although care must be taken that it is added to the second, and not the first, Piola-Kirchhoff stress tensor \cite{destrade13}.

In terms of the Cauchy stress tensor, the viscous stress \eqref{GL} reads $\bm{\sigma}^\text{v} = 2\eta \bm{B} {\bm D}{\bm B}$, where we have used the identity $\dot{\bm E} = \bm{F}^\text{T} \bm{D} \bm{F}$ and the definition $\bm{B} = \bm{F}\bm{F}^\text{T}$ of the left Cauchy-Green deformation tensor. Here, we note that the viscous Cauchy stress is not necessarily trace-free (the Cauchy-Schwarz inequality does not apply). Nevertheless, the latter can still be replaced by its deviatoric part $\bm{\sigma}^\text{v}_\text{d}$ up to a suitable redefinition of the arbitrary Lagrange multiplier in Eq.~\eqref{Constitutive}.

Furthermore, it is worth pointing out that the viscosity theories \eqref{Newtonian}-\eqref{GL} have fundamentally different mathematical properties. In fact, contrary to the Newtonian viscosity case \eqref{Newtonian}, viscoelastic shearing motions might be limited to a finite time when using \eqref{GL}, which is potentially problematic for the purpose of experimental characterisation involving long-time relaxation processes \cite{berjamin23}.

In the next section, we present straightforward generalisations of the viscous stresses \eqref{Newtonian}-\eqref{GL} to fractional orders \eqref{Frac}, and establish connections with the literature (Section \ref{sec:Model}). We briefly discuss approximations of the fractional derivative to be used in computational applications (Section \ref{sec:Frac}). Then, we discuss the physical properties of the models at hand, more specifically in relation with the objectivity requirement and thermodynamic consistency (Section \ref{sec:Prop}). It appears that some theories from the literature are physically unacceptable in those respects. Furthermore, we study elementary shearing and tensile motions (Section \ref{sec:Elem}). Finally, we compute incremental stresses via a `small-on-large' linearisation and obtain acoustoelastic formulas (Section \ref{sec:SoL}). The results of this study might be used in experimental setups or in other applications.


\section{Constitutive models}\label{sec:Model}


In this section, we introduce straightforward generalisations of the viscous stresses \eqref{Newtonian}-\eqref{GL} to fractional differential orders \eqref{Frac}. A summary of these constitutive laws is given in Table~\ref{tab:Constit}.

\begin{table}
	\centering
	\caption{Expression of the viscous Cauchy stress $\bm{\sigma}^\text{v} = \bm{F} \bm{S}^\text{v} \bm{F}^\text{T}$ for the Models A, B, C, including the elastic limit $\alpha\to 0$ and the viscous limit $\alpha\to 1$. The fractional time derivative $\mathfrak{D}^\alpha$ is defined in \eqref{Frac} and the fractional viscosity $\eta_\alpha$ is expressed in $\text{Pa.s}^\alpha$. \label{tab:Constit}}
	\renewcommand{\arraystretch}{1.1}
		\begin{tabular}{ccccc}
			\toprule
			\multicolumn{2}{c}{ Model } & Fractional viscous stress $\bm{\sigma}^\text{v}$ & Elastic limit & Viscous limit \\
			\midrule
			A & \eqref{Shen} & $2\eta_\alpha \bm{F}(\mathfrak{D}^\alpha {\bm \Pi})\bm{F}^\text{T}$ & $ \mu (\bm{B} - \bm{I})$ & $2\eta \bm{D}$ \\
			B & \eqref{HauptBCauchy} & $2\eta_\alpha \bm{F}^{-\text{T}}(\mathfrak{D}^\alpha {\bm E})\bm{F}^{-1}$ & $\mu (\bm{I} - \bm{B}^{-1})$ & $2\eta \bm{D}$ \\
			C & \eqref{Capilnasiu} & $2\eta_\alpha \bm{F}(\mathfrak{D}^\alpha {\bm E})\bm{F}^\text{T}$ & $ \mu ( \bm{B}^2 - \bm{B})$ & $2\eta \bm{B}{\bm D}\bm{B}$ \\
			\bottomrule
	\end{tabular}
\end{table}

\subsection{Model A}

We introduce a fractional time derivative \eqref{Frac} of the Piola strain ${\bm \Pi}$ in the definition of the Newtonian viscous Piola-Kirchhoff stress  \eqref{Newtonian}, as follows: $\bm{S}^\text{v} = 2\eta_{\alpha} \, \mathfrak{D}^\alpha {\bm \Pi}$, where $\eta_{\alpha}>0$ is a fractional dynamic viscosity (in Pa.s\textsuperscript{$\alpha$}). This expression involves only two independent parameters, the coefficient $\eta_{\alpha}$ and the differential order $\alpha$. For convenience, we introduce the redundant parameter $\mu = \eta_\alpha/\tau^\alpha$ where $\tau>0$ is a characteristic time. In other words, we have $\bm{S}^\text{v} = 2\mu \tau^{\alpha} \, \mathfrak{D}^\alpha {\bm \Pi}$. Without loss of generality, the parameter $\mu$ can be chosen in such a way that it equals the initial shear modulus in solid materials (in Pa).

According to the definition of the fractional derivative \eqref{Frac}, this expression might be rewritten as a time-domain convolution product with kernel $\kappa$,
\begin{equation}
	\bm{S}^\text{v} = 2\mu\, \kappa * \dot{\bm \Pi},
	\qquad \text{where} \qquad
	\kappa(t) = \frac{(t/\tau)^{-\alpha}}{\Gamma (1-\alpha)} \text{H}(t) ,
	\label{Kernel}
\end{equation}
and $\text{H}$ represents the Heaviside step function. In agreement with previous definitions, we thus have the relationship $\kappa * \dot{\bm \Pi} = \tau^\alpha\, \mathfrak{D}^\alpha {\bm \Pi}$. In terms of the Cauchy stress tensor, we find
\begin{equation}
	\bm{\sigma}^\text{v} = 2\mu\, \bm{F} (\kappa * \dot{\bm \Pi}) \bm{F}^\text{T} = 2\mu\, \kappa * (\bm{F}_{t|s}{\bm D}\bm{F}_{t|s}^\text{T}) ,
	\label{Shen}
\end{equation}
due to the identity $\dot{\bm \Pi} = \bm{F}^{-1}{\bm D}\bm{F}^{-\text{T}}$. Here, the tensor
\begin{equation}
\bm{F}_{t|s} = \bm{F}(t) \bm{F}^{-1}(s)
\label{FRel}
\end{equation}
is the relative deformation gradient from the configuration at the integration time $s \in \left[0,t\right]$ to the configuration at the current time $t$. We emphasize that other expressions for the convolution kernel are possible, see the review by Freed and Diethelm \cite{freed06} for alternatives.

With the above expressions of the viscous stress, we recover the Newtonian viscous stress $\bm{S}^\text{v} \to 2\eta \dot{\bm \Pi}$, i.e. $\bm{\sigma}^\text{v} \to 2\eta \bm D$, in the limit of integer differentiation $\alpha \to 1$ where $\eta=\mu \tau$. In the limit of no differentiation $\alpha \to 0$, we obtain the extra elastic contribution $\bm{S}^\text{v} \to 2\mu {\bm \Pi}$, i.e. $\bm{\sigma}^\text{v} \to \mu (\bm B - \bm I)$, which is of neo-Hookean type{\,---\,}this property can be inferred from a redefinition of the Lagrange multiplier in Eq.~\eqref{Constitutive}\textsubscript{b}, see also the expression \eqref{Mooney} of the Mooney-Rivlin stress with the second Mooney parameter $\mathfrak{C}_-$ equal to zero. Therefore, this model manages to cover and connect an elastic theory and a viscous theory.

The present theory is strongly related to other models found in the literature. Indeed, Eq.~\eqref{Shen} matches Eq.~(2.43) of Shen \cite{shen20}, where it is linked to the theory by Drozdov \cite{drozdov97} (cf. next paragraph). It is also of the general form proposed in Capilnasiu et al. \cite{capilnasiu20}, Eq.~(5) therein, although it does not match later propositions from that study. Furthermore, the above expression is found in Palade et al.~\cite{palade99} as a special case of Eq.~(16) therein. It is also aligned with `Model~A' of Haupt and Lion \cite{haupt02}, see Eq.~(5.12) therein. In particular, if the kernel $\kappa$ is chosen exponential instead of the power-law expression \eqref{Kernel}, then we recover the `upper-convected' Maxwell model which involves the Oldroyd rate of Cauchy stress (also equivalent to the Truesdell rate in the incompressible case).

Eqs.~(27)-(32) of Drozdov \cite{drozdov97} introduce a viscous stress based on a relative deformation gradient tensor and the strain-rate tensor $\bm D$. However, the conventions therein differ from the present ones. To reconcile the two, we take the transpose of the deformation gradients in \cite{drozdov97}, see also Eq.~(4) and later sections therein. Thus, Eq.~(3) of \cite{drozdov97} becomes $\bm{F}_{t|s}^\text{T} = \bm{F}^{-\text{T}}(s) \bm{F}^\text{T}(t)$ if our notations \eqref{FRel} are used, and the viscous stress (32) proposed therein takes the form of Eq.~\eqref{Shen}\textsubscript{b}.

\subsection{Model B}

We introduce a fractional time derivative \eqref{Frac} of the Green-Lagrange strain ${\bm E}$ in the definition of the viscous stress \eqref{Newtonian} as follows: $\bm{S}^\text{v} = 2\eta_{\alpha} \bm{C}^{-1} (\mathfrak{D}^\alpha {\bm E}) \bm{C}^{-1}$, i.e.
\begin{align}
	&\bm{S}^\text{v} = 2\mu\, \bm{C}^{-1} (\kappa * \dot{\bm E}) \bm{C}^{-1} ,
	\label{HauptB} \\
	&\bm{\sigma}^\text{v} = 2\mu\, \bm{F}^{-\text{T}}(\kappa * \dot{\bm E}) \bm{F}^{-1} = 2\mu\, \kappa * (\bm{F}_{t|s}^{-\text{T}}{\bm D}\bm{F}_{t|s}^{-1}) , \label{HauptBCauchy}
\end{align}
where we have used the same notations as for Model~A, as well as the identity $\dot{\bm E} =  \bm{F}^\text{T} \bm{D} \bm{F}$. Thus, the Cauchy stress has a similar expression as in \eqref{Shen}, up to the fact that the relative deformation gradient tensor $\bm{F}_{t|s}$ has been replaced with its inverse transpose.

With the above expression, we recover the same viscous limit as for Model~A when $\alpha \to 1$. In the limit of no differentiation $\alpha \to 0$, Eq.~\eqref{HauptB} yields the elastic stress $\bm{S}^\text{v} \to 2\mu \bm{\Pi}\bm{C}^{-1}$, i.e. $\bm{\sigma}^\text{v} \to \mu (\bm{I}-\bm{B}^{-1})$. Thus, up to a redefinition of the Lagrange multiplier in \eqref{Constitutive}\textsubscript{b}, we observe that the elastic limit of this theory corresponds to Mooney-Rivlin elasticity \eqref{Mooney} with the first Mooney parameter $\mathfrak{C}_+$ equal to zero. The viscous stress \eqref{HauptB} corresponds to Model~B of the study by Haupt and Lion \cite{haupt02}, see Eq.~(5.22) therein. In particular, if the kernel $\kappa$ is chosen exponential instead of the power-law expression \eqref{Kernel}, then we recover the `lower-convected' Maxwell model involving the Cotter-Rivlin rate of Cauchy stress.

\subsection{Model C}

We introduce a fractional time derivative \eqref{Frac} of the Green-Lagrange strain ${\bm E}$ in the definition of the second form of viscous stress \eqref{GL} as follows: $\bm{S}^\text{v} = 2\eta_{\alpha} \, \mathfrak{D}^\alpha {\bm E}$, i.e.
\begin{equation}
	\bm{S}^\text{v} = 2\mu\, \kappa * \dot{\bm E} ,
	\qquad
	\bm{\sigma}^\text{v} = 2\mu\, \bm{F}(\kappa * \dot{\bm E})\bm{F}^\text{T} ,
	\label{Capilnasiu}
\end{equation}
where we have used the same notations as for Model~A. The only difference with respect to Model~B is the multiplication of the second Piola-Kirchhoff stress tensor by $\bm{C}^{-1}$ on the left and on the right in the latter case.

With the above expression, we recover $\bm{S}^\text{v} \to 2\eta \dot{\bm E}$, i.e. $\bm{\sigma}^\text{v} \to 2\eta \bm{BDB}$, in the limit of integer differentiation $\alpha \to 1$. 
In the limit of no differentiation $\alpha \to 0$, Eq.~\eqref{Capilnasiu} yields the elastic stress $\bm{S}^\text{v} \to 2\mu \bm{E}$, i.e. $\bm{\sigma}^\text{v} \to \mu (\bm B^2 - \bm B)$. We note that the proposed viscous stress is of the general form found in Capilnasiu et al. \cite{capilnasiu20}. Using the identity $\dot{\bm E} =  \bm{F}^\text{T} \bm{D} \bm{F}$, we observe that the viscous stress \eqref{Capilnasiu} is included in Eq.~(18) of Palade et al.~\cite{palade99} as a special case. In particular, if the kernel $\kappa$ is chosen exponential instead of the power-law expression \eqref{Kernel}, then we recover a Maxwell-type model involving the material rate of second Piola-Kirchhoff stress. A scalar compressible model of this form was also proposed by Sugimoto \cite{sugimoto89}, Eq.~(3.7) therein.

\subsection{Other models}

To facilitate comparisons, the viscous stresses for the Models A, B, C are summarised in Table~\ref{tab:Constit}. Alternative modelling approaches can lead to different definitions of the fractional viscous stress. Below, we list other propositions found in literature.
\begin{itemize}
	\item Freed and Diethelm \cite{freed06} propose to exploit the K-BKZ hypothesis \cite{kaye62,bernstein63} (after Kaye, Bernstein, Kearsley, Zapas), which consists of a viscoelastic extension of elastic constitutive models based initially on exponential relaxation kernels. Variations of this theory are provided by Coleman and Noll \cite{coleman61} (see Eq.~(5.20) therein). Based on more general expressions of the stored viscous energy than for Model~A, see Rao and Rajagopal \cite{rao07b}, these models can lead to rather complex expressions of the fractional viscous stress.
	\item Another theory by Freed and Diethelm \cite{freed07} includes formally similar viscous stresses to Eq.~\eqref{Shen}\textsubscript{b} up to the substitution of the relative deformation gradient $\bm{F}_{t|s}$ through the relative rotation $\bm{R}_{t|s} = \bm{R}(t) \bm{R}^{\text{T}}(s)$, where $\bm R$ is the proper orthogonal tensor in the polar decomposition of $\bm F$ (see e.g. Holzapfel~\cite{holzapfel00}).
	\item A given elastic stress $\bm{S}^\text{r}$ can be used to express the fractional viscous stress as $\bm{S}^\text{v} = \kappa * \dot{{\bm S}^\text{r}_\text{D}}$, which involves the material time derivative of the deviator ${\bm S}^\text{r}_\text{D}$, see \cite{capilnasiu20,zhang20}. This way, we have the limit $\bm{S}^\text{v} \to {\bm S}^\text{r}_\text{D}$ as $\alpha\to 0$. Here, the elastic stress $\bm{S}^\text{r}$ is not necessarily identical to the elastic response $\bm{S}^\text{e}$. For instance, one might set $\bm{S}^\text{r} = \mu^\text{r}\bm{I}$ if this elastic stress is chosen neo-Hookean, where $\mu^\text{r}$ is a given shear modulus. The present approach is further investigated in a recent preprint \cite{jiang23}.
	\item Based on the same definition of the relative deformation gradient tensor $\bm{F}_{t|s}$ as in Eq.~\eqref{FRel}, Zhao et al. \cite{zhao21} switch the position of the transposition symbols in Eq.~\eqref{Shen}\textsubscript{b} to mimic Ref.~\cite{drozdov97} (see Eqs.~(2)-(10) of Ref. \cite{zhao21} and also Eqs.~(6)-(8) of Gao et al. \cite{gao23}).
	\item Delory et al. \cite{delory23} introduce pseudo-Newtonian stresses based on a fractional rate of deformation gradient $\dot{\bm F}$, invoking Zhang et al. \cite{zhang22} where fractional viscosity is incorporated in a linear fashion.
\end{itemize}

\section{Approximation of the fractional derivative}\label{sec:Frac}

For practical use, we represent the convolution kernel of Eq.~\eqref{Kernel} by a continuous sum of exponentials
\begin{equation}
	\kappa(t) = \text{H}(t) \int_0^\infty \phi(\zeta)\, \text{e}^{-t/\zeta} \text{d}\zeta ,
	\qquad
	\phi(\zeta) = \frac{\tau^\alpha \sin(\alpha\pi)}{\pi\, \zeta^{\alpha+1}} ,
	\label{Spectrum}
\end{equation}
with a suitable expression of the spectrum $\phi$, see Lion~\cite{lion97} and Euler's reflection formula. This particular form of $\kappa$ is a \emph{diffusive representation} of the fractional derivative, see Section~7.4.1 of Matignon~\cite{matignon09} where the same expression is proposed up to a change of variable in the integral.

In practice, the continuous spectrum of relaxation \eqref{Spectrum} might be approximated by a discrete one, which leads to a Prony-type theory with exponential relaxation \cite{zhang20}. A straightforward discretisation of this kind is obtained based on Gauss-Laguerre quadrature, which consists in evaluating the integrand of \eqref{Spectrum} at the roots of the Laguerre polynomial of degree $N$ with suitable weights. Faster convergence can be obtained based on a change of variable followed by Gauss-Jacobi quadrature, see Diethelm \cite{diethelm09} as well as Birk and Song \cite{birk10} for more elaborated techniques.

For the purpose of illustration, we consider a straightforward Prony series approximation here, and we discuss its suitability. Thus, we write
\begin{equation}
	\kappa(t) = \frac{\text{sin}(\pi\alpha)}{\pi\alpha}\, \text{H}(t) \int_0^1 \frac{\text{e}^{-t/\zeta}}{\theta} \text{d}\theta \simeq \kappa_N(t) ,
	\label{Quad}
\end{equation}
where
\begin{equation}
	 \kappa_N(t) = \text{H}(t) \sum_{n=1}^N w_n \text{e}^{-t/\zeta_n} ,\qquad w_n = \frac{\text{sin}(\pi\alpha)}{\pi\alpha N \theta_n} .
	\label{QuadProny}
\end{equation}
Here, we have used the change of variables $\zeta = \tau\, (-\ln \theta)^{-1/\alpha}$ in the integral \eqref{Spectrum} over the time coordinate $\zeta$. The resulting integral over $\theta = \exp(-(\zeta/\tau)^{-\alpha})$ was then approximated as a discrete sum based on the extended midpoint rule with $N$ points $\theta_n = \frac{2n-1}{2N}$ for $1\leq n\leq N$, which correspond to the relaxation times $\zeta_n = \zeta|_{\theta=\theta_n}$.

\begin{figure}
	\begin{minipage}{0.49\textwidth}
		\centering
		(a)
		
		\includegraphics{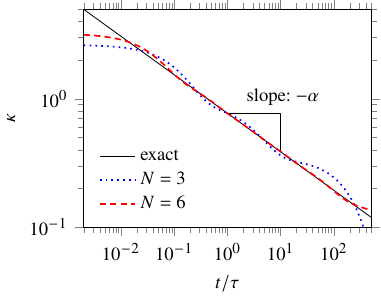}
	\end{minipage}\hfill
	\begin{minipage}{0.49\textwidth}
		\centering
		(b)
		
		\includegraphics{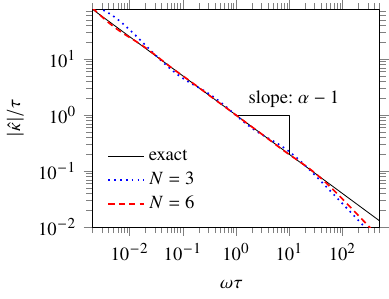}
	\end{minipage}
	
	\caption{Exact convolution kernel (full line) and its diffusive Prony series approximation for $\alpha = 0.3$ with $N=3$ (dotted line) and $N=6$ terms (dashed lines). (a) Impulse response, (b) Fourier spectrum. \label{fig:Kernel}}
\end{figure}

Such an approximation of the kernel $\kappa$ as a finite sum of exponentials is rather accurate over a broad range of times and frequencies provided a sufficient number $N$ of relaxation mechanisms is included, see the example in Figure \ref{fig:Kernel} where we have used $N=3$ and $N=6$. Therein, we display also the modulus of the Fourier transform $\hat \kappa(\omega) = \int \kappa(t) \text{e}^{-\text{i}\omega t} \text{d}t$ of $\kappa$ and $\kappa_N$, which satisfies
\begin{equation}
	\hat \kappa(\omega)/\tau = (\text{i}\omega \tau)^{\alpha-1} , \qquad \hat \kappa_N(\omega) = \sum_{n=1}^N \frac{w_n \zeta_n}{1+\text{i}\omega \zeta_n} .
\end{equation}
In particular, the figure illustrates the error introduced by the bounded Prony series approximation \eqref{QuadProny} at short times where $\kappa$ is singular. Nevertheless, the figure shows that the Prony series approximation obtained for $N=6$ relaxation mechanisms is much more accurate than that obtained for $N=3$ relaxation mechanisms at long times, despite the singularity of $\hat\kappa$ at low frequency.

Based on a representation of the form \eqref{Spectrum}-\eqref{Quad}, the following identity holds for any causal tensor field $\bm T$,
\begin{equation}
	\kappa * \dot{\bm T} = \int_0^\infty \phi(\zeta)\, \bm{T}_{\zeta}^\text{v} \, \text{d}\zeta \simeq \sum_{n=1}^N  w_n \bm{T}_{\zeta_n}^\text{v} .
	\label{QuadApprox}
\end{equation}
The tensors
\begin{equation}
	\bm{T}_{\zeta}^\text{v} = \int_0^t \text{e}^{-(t-s)/\zeta} \dot{\bm T}(s)\, \text{d}s
	\label{QuadMem}
\end{equation}
are memory variables governed by a differential equation of the form
\begin{equation}
	\dot{\bm T}_{\zeta}^\text{v} = \dot{\bm T} - {\bm T}_{\zeta}^\text{v}/\zeta .
	\label{QuadDiff}
\end{equation}
This result can be obtained by application of the Leibniz integral rule to Eq.~\eqref{QuadMem}\textsubscript{b}. Under this form, evaluation of the fractional derivative amounts to solving a linear system of differential equations.

Such approximations of the fractional derivative are therefore convenient from a computational point of view as they avoid the storage of the history of $\bm T$ to evaluate the current viscous stresses. Nevertheless, it is worth pointing out that special care should be taken as the system \eqref{QuadDiff} might become stiff. In fact, if we consider the approximation \eqref{QuadProny} at large $N$, the smallest dimensionless relaxation time $\zeta_1/\tau = \ln(2N)^{-1/\alpha}$ approaches zero, whereas the largest relaxation time $\zeta_N/\tau = \ln(\frac{2N}{2N-1})^{-1/\alpha}$ approaches infinity. Further improvements of the approximation \eqref{QuadProny} can be obtained through optimisation of the $2N$ coefficients $\zeta_n$, $w_n$, see for instance Blanc et al. \cite{blanc16}.


\section{Properties}\label{sec:Prop}



\subsection{Material frame-indifference}\label{subsec:Obs}


Here we examine the acceptability of the above-mentioned models in terms of the material frame-indifference principle, which stipulates that the mechanical response of a material should not be affected by a change of observer (Holzapfel \cite{holzapfel00}, Sections~5.2-5.4). Doing so, we show that some of the presented models do not comply with the material frame-indifference principle.

We consider the superimposed rigid-body motion defined by $\bm{x}^+ = \bm{c} + \bm{Q} \bm{x}$ where $\bm{c}(t)$ is a vector and $\bm{Q}(t)$ is a proper orthogonal tensor. 
Then the deformation gradient tensor transforms according to $\bm{F}^+ = \bm{Q}\bm{F}$. 
The kinematic variables $\bm{\Pi}$, $\bm{E}$ and their material time-derivatives are unaffected by the rigid-body motion: $\bm{\Pi} = \bm{\Pi}^+$, $\bm{E} = \bm{E}^+$, as are second Piola-Kirchhoff stresses: $\bm{S}^+ = \bm{S}$. 
Using the product rule, we derive the change of observer formula: $\dot{\bm F}^+ = \dot{\bm Q}\bm{F} + \bm{Q}\dot{\bm F}$, leading to: $\bm{L}^+ = \bm{\Omega} + \bm{Q}\bm{L}\bm{Q}^\text{T}$, where $\bm{\Omega} = \dot{\bm Q}\bm{Q}^\text{T}$ is skew-symmetric, showing that $\bm L$ is not an objective tensor. 
The Eulerian strain rate tensor and the Cauchy stress tensor are objective, because they satisfy the change of observer formulas $\bm{D}^+ = \bm{Q}\bm{D}\bm{Q}^\text{T}$ and $\bm{\sigma}^+ = \bm{Q}\bm{\sigma}\bm{Q}^\text{T}$.

Now introduce the polar decomposition of the deformation gradient tensor: $\bm{F} = \bm{R}\bm{U} = \bm{V}\bm{R}$ where the stretch tensors $\bm{U}$, $\bm{V}$ are positive definite and symmetric. The tensor $\bm R$ is a proper orthogonal tensor which satisfies the transformation rule $\bm{R}^+ = \bm{Q}\bm{R}$. Based on the definition of the relative deformation gradient tensor $\bm{F}_{t|s}$ in \eqref{FRel} and of the relative rotation tensor $\bm{R}_{t|s}$, we derive the following transformation rules for these quantities: $\bm{F}_{t|s}^+ = \bm{Q}(t)\bm{F}_{t|s}\bm{Q}^\text{T}(s)$ and $\bm{R}_{t|s}^+ = \bm{Q}(t)\bm{R}_{t|s}\bm{Q}^\text{T}(s)$.

The viscous Piola-Kirchhoff stress $\bm{S}^\text{v}$ defined in Eqs.~\eqref{Kernel}-\eqref{HauptB}-\eqref{Capilnasiu} must remain invariant. 
Because  $\dot{\bm\Pi}$ and $\dot{\bm E}$ are unaffected by the superimposed rigid-body motion, the present constitutive laws are frame-indifferent. An alternative proof for Models~A and B is to use the transformation rule for $\bm{F}_{t|s}$ to show that Eq.~\eqref{Shen}\textsubscript{b} and \eqref{HauptBCauchy}\textsubscript{b} are frame-indifferent.

\begin{remark}
Using the transformation rule for the relative deformation gradient tensor, one shows that the K-BKZ viscous stress is frame-indifferent. The transformation rule for $\bm{R}_{t|s}$ yields the acceptability of the theory by Freed and Diethelm \cite{freed07} from the point of view of material frame-indifference. Given that the models found in \cite{capilnasiu20,zhang20} involve only invariant quantities (e.g., second Piola-Kirchhoff stresses and their material rates), the material frame-indifference property is straightforwardly satisfied for these theories.
\end{remark}

\begin{remark}
Zhao et al. \cite{zhao21} propose $\bm{\sigma}^{\text{v}} = 2\mu\, \kappa * (\bm{F}_{t|s}^\text{T}{\bm D}\bm{F}_{t|s})$. Then, superimposition of the rigid-body motion $\bm{x}^+ = \bm{c} + \bm{Q} \bm{x}$ provides
\begin{equation}
	\bm{\sigma}^{\text{v}+} = 2\mu\, \bm{Q}\big( \kappa * (\bm{\Theta}^\text{T}\bm{F}^\text{T}_{t|s}\bm{\Theta}^\text{T}{\bm D}\, \bm{\Theta}\bm{F}_{t|s}\bm{\Theta})\big) \bm{Q}^\text{T} , \qquad \bm{\Theta} = \bm{Q}^{\text{T}}(s)\bm{Q}(t) .
\end{equation}
Delory et al. \cite{delory23} propose $\bm{\sigma}^\text{v} = \mu\, (\bm{l} + \bm{l}^{\text{T}})$ with $\bm{l} = (\kappa * \dot{\bm F})\bm{F}^{-1}$. This way,
\begin{equation}
	\begin{aligned}
		\bm{\sigma}^{\text{v}+} &= \mu\, \bm{Q} \big( \kappa * (\bm{\Theta}^\text{T}\bm{Q}^\text{T}\dot{\bm Q}\bm{F}_{t|s}^{-1} - \bm{F}_{t|s}^{-\text{T}} \bm{Q}^\text{T}\dot{\bm Q}\bm{\Theta}) \big) \bm{Q}^\text{T} \\
		&\quad  + \mu\, \bm{Q} \big( (\kappa * \bm{\Theta}^\text{T}\dot{\bm F}) \bm{F}^{-1} + \bm{F}^{-\text{T}} (\kappa * \dot{\bm F}^\text{T}\bm{\Theta}) \big) \bm{Q}^\text{T} ,
	\end{aligned}
\end{equation}
where the skew-symmetry of $\bm{\Omega}$ was used. As shown in the above computations, we note that  $\bm{\sigma}^{\text{v}+} \neq \bm{Q}\bm{\sigma}^\text{v}\bm{Q}^\text{T}$ in general, for both theories. Therefore, these constitutive laws are not frame-indifferent.
\end{remark}


\subsection{Thermodynamic consistency}\label{subsec:Thermo}


Thermodynamic consistency of Models~A and B was proved by Haupt and Lion \cite{haupt02}. For Model~C, thermodynamic consistency can be obtained in a similar way to the linear case \cite{lion97}. To do so, we use the diffusive representation \eqref{Spectrum} to rewrite the viscous stress \eqref{Capilnasiu}\textsubscript{a} as $\bm{S}^\text{v} = 2\mu \int \phi(\zeta) \bm{E}^\text{v}_\zeta\, \text{d}\zeta$ by reversing the order of integration, where $\bm{E}^\text{v}_\zeta$ is a memory variable defined in Eq.~\eqref{QuadMem} with ${\bm T} = {\bm E}$.

In an isothermal modelling framework, the free energy per unit of reference volume is then defined as
\begin{equation}
\Psi = \Psi^\text{e} + \Psi^\text{v} ,
\label{Thermo}
\end{equation}
where $\Psi^\text{e}$ is the strain energy function associated with the elastic stress contribution, $\Psi^\text{v} = \mu \int_0^\infty \phi(\zeta) \|\bm{E}^\text{v}_\zeta\|^2 \text{d}\zeta$ corresponds to the viscous part, and $\| \cdot \|$ is the Frobenius norm. According to the first and second principles of thermodynamics, the dissipation $\mathscr{D} = \bm{S}:\dot{\bm E} - \dot{\Psi}$ per unit of reference volume must remain non-negative at all times, where the colon symbol denotes double contraction, aka the Frobenius inner product. Using the above expression of $\Psi$, the inequality $\mathscr{D}\geq 0$ is obtained straightforwardly (see Section~6.3 of Holzapfel \cite{holzapfel00} for the inclusion of incompressibility).

The dissipative behaviour of the discretised version \eqref{QuadProny} of this theory follows immediately. In fact, it suffices to define the thermodynamic potential $\Psi^\text{v} = \mu \sum_n w_n \| \bm{E}^\text{v}_{\zeta_n}\|^2$ for the viscous part along with the stress $\bm{S}^\text{v} = 2\mu \sum_n w_n \bm{E}^\text{v}_{\zeta_n}$. Under this form, similarities with Prony series theories from the literature can be identified \cite{berjamin22}.

\begin{remark}
The thermodynamic admissibility of the K-BKZ theory was studied by Bernstein et al. \cite{bernstein64}, see also Rao and Rajagopal \cite{rao07}. Freed and Diethelm \cite{freed07} leave thermodynamic consistency to the reader's curiosity, and Capilnasiu et al. \cite{capilnasiu20,zhang20} do not provide the thermodynamic potentials related to their model either.
\end{remark}


\section{Elementary motions}\label{sec:Elem}


Various illustrations are provided in the following sections, including simple shear and uniaxial tensile motions. The Cauchy stress tensor $\bm \sigma$ is deduced from Eq.~\eqref{Constitutive}\textsubscript{b} with suitable constitutive assumptions for the elastic and viscous parts. Here, we assume that the elastic response is of Mooney-Rivlin type, i.e.,
\begin{equation}
	\bm{\sigma}^\text{e} = 2\mathfrak{C}_+ \bm{B} - 2\mathfrak{C}_- \bm{B}^{-1} , \qquad 2\mathfrak{C}_\pm = \tfrac12\mu (1\pm\beta) ,
	\label{Mooney}
\end{equation}
where $\bm{B} = \bm{F}\bm{F}^\text{T}$ is the right Cauchy-Green strain tensor, $\mu > 0$ is the shear modulus, and the parameters $\mathfrak{C}_\pm$ are the Mooney coefficients. The parameter ${-1} \leq \beta \leq 1$ is introduced in such a way that $\beta=1$ entails neo-Hookean material behaviour. For sake of clarity, we now reduce the discussion to Models A, B and C. Thus, the viscous stress $\bm{\sigma}^\text{v}$ is deduced from Eqs.~\eqref{Shen}-\eqref{HauptBCauchy}-\eqref{Capilnasiu}, respectively, see also the expressions in Table~\ref{tab:Constit}.

\subsection{Simple shear}

We consider general simple shear motions described by the deformation gradient tensor $\bm{F} = \bm{I} + \gamma\, (\bm{e}_x \otimes \bm{e}_z)$, where $\gamma(z,t)$ is the shear strain, and the vectors $\bm{e}_x, \bm{e}_y, \bm{e}_z$ form an orthonormal basis. With the present assumptions, we obtain the corresponding stress-strain relationships
\begin{equation}
	\Sigma = \epsilon + \tau^\alpha\, \mathfrak{D}^\alpha\epsilon ,\qquad
	\Sigma = \epsilon + \tau^\alpha\, ( \mathfrak{D}^\alpha \epsilon + K^2 \epsilon\, \mathfrak{D}^\alpha \epsilon^2) ,
	\label{Shear}
\end{equation}
for Models~A-B and for Model~C, which correspond to Eqs.~\eqref{Shear}\textsubscript{a} and \eqref{Shear}\textsubscript{b} respectively.
Here, $\Sigma = \sigma_{13}/(\mu K)$ is a non-dimensional measure of the shear stress, $\epsilon = \gamma/K$ is a rescaled shear strain, and $K>0$ is a given strain magnitude. We observe that Models A and B lead to a linear fractional Kelvin-Voigt rheology. Note in passing that Models A-B and C produce the same expression of the shear stress in the limit of infinitesimal shear strains $K\to 0$, namely \eqref{Shear}\textsubscript{a}.

\subsubsection{Shear creep.}\label{subsubsec:Creep}

In a standard fashion \cite{carcione15}, we now assume that the material is initially at rest, and that it is suddenly subjected to a step shear stress $\Sigma = \text{H}(t)$ which
entails a simple shear deformation. The evolution of the strain is governed by the fractional differential equations resulting from the above constitutive relationships.

\paragraph{Models A, B.}

\begin{figure}
	\begin{minipage}{0.49\textwidth}
		\centering
		(a)
		
		\includegraphics{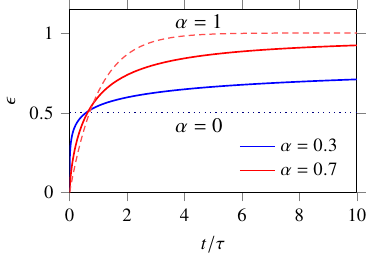}
	\end{minipage}\hfill
	\begin{minipage}{0.49\textwidth}
		\centering
		(b)
		
		\includegraphics{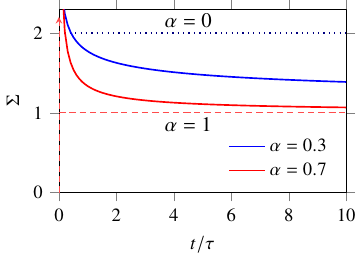}
	\end{minipage}
	\caption{Time-dependent shearing motion for Models~A and B. (a) Creep function, i.e., evolution of the strain for a constant applied stress; (b) Relaxation function, i.e., evolution of the stress for a constant applied strain. \label{fig:Creep}}
\end{figure}

In this case, the relationship \eqref{Shear}\textsubscript{a} with $\Sigma = \text{H}$ leads to the linear fractional differential equation $\epsilon + \tau^\alpha\, \mathfrak{D}^\alpha\epsilon = \text{H}$. Solutions are given by the creep function \cite{mainardi11}
\begin{equation}
	\epsilon(t) = 1 - \text{E}_\alpha(-(t/\tau)^\alpha) , \qquad \text{E}_\alpha(z) = \sum_{n=0}^\infty \frac{z^n}{\Gamma(\alpha n+ 1)} ,
	\label{Mittag}
\end{equation}
where $\text{E}_\alpha$ is the one-parameter Mittag-Leffler function, see also Ref. \cite{podlubny99}. Illustrations are provided in Figure~\ref{fig:Creep}a for several values of $\alpha$. In the elastic limit $\alpha\to 0$, the creep response is a step shear strain (dotted line), whereas in the viscous limit $\alpha\to 1$, we recover an exponential creep response \cite{carcione15} (dashed line).

\paragraph{Model C.}

Here, we obtain a nonlinear fractional differential equation deduced from \eqref{Shear}\textsubscript{b} with $\Sigma = \text{H}$.
In the present case, exact resolution of the creep problem seems hardly tractable. Approximate resolution might be performed, for instance based on the representation \eqref{Spectrum}-\eqref{Quad} of the fractional derivative, or on a perturbation approach involving the small parameter $K$, see for instance \cite{pucci15}.

\subsubsection{Shear stress relaxation.}\label{subsec:Relax}

Initially at rest, the material is suddenly subjected to a step shear strain $\epsilon = \text{H}(t)$. The evolution of the stress is deduced from Eq.~\eqref{Shear}.

\paragraph{Models A, B.}

The stress-strain relationship \eqref{Shear}\textsubscript{a} produces $\Sigma = \epsilon + \kappa * \dot{\epsilon}$, where we have rewritten the fractional derivative \eqref{Frac} in the form of a convolution product.
At positive times, we therefore find $\Sigma = 1 + \kappa$ where the kernel $\kappa(t)$ in Eq.~\eqref{Kernel} follows a power-law evolution. Note in passing that the first (unit) term vanishes in fluid materials. Illustrations are provided in Figures~\ref{fig:Kernel} and \ref{fig:Creep}b, which can be compared to experimental results from the literature, see Figure~\ref{fig-Bonfanti}. In the elastic limit $\alpha\to 0$, the relaxation response is a step shear stress (dotted line), whereas in the viscous limit $\alpha\to 1$, the relaxation response is singular (dashed line). Unlike their fractional counterpart, classical Kelvin-Voigt models cannot account for stress relaxation \cite{carcione15}.

\paragraph{Model C.}

Here, direct evaluation of the stress is not straightforward (the nonlinear term of \eqref{Shear}\textsubscript{b} with $\epsilon = \text{H}$ seems not well-defined). Nevertheless, noting that $\text{H}^2 = \text{H}$ in the weak sense, one would obtain $\Sigma = 1 + (1+K^2)\,\kappa$ at positive times, provided that every step of this computation is mathematically meaningful. This way, the curves displayed in Figure \ref{fig:Creep}b undergo a vertical dilation as $K$ is increased.


\subsection{Uniaxial tension-compression}\label{subsec:Tens}


We consider a state of uniaxial tension described by the diagonal deformation gradient tensor $\bm{F} = \text{diag}[\lambda , \lambda^{-1/2}, \lambda^{-1/2}]$ at all times. Consequently, the Cauchy stress tensor $\bm \sigma$ is diagonal and its lateral components are equal, $\sigma_{22} = \sigma_{33}$. By making these lateral stresses vanish, the constitutive law \eqref{Constitutive} yields an additive decomposition of the tensile Cauchy stress,
\begin{equation}
	\sigma_{11} = \sigma_{11}^\text{e} + \sigma_{11}^\text{v}, 
\end{equation}
where the elastic part $\sigma_{11}^\text{e} = 2 \left(\lambda \mathfrak{C}_+ + \mathfrak{C}_-\right) (\lambda -\lambda^{-2})$ follows from the Mooney-Rivlin model \eqref{Mooney}. The tensile component of the viscous stress deduced from Models A, B, C satisfies
\begin{equation}
	\begin{aligned}
	\sigma_{11}^\text{v}/\eta_\alpha &= \lambda^{-1}\, \mathfrak{D}^\alpha \lambda - \lambda^{2}\, \mathfrak{D}^\alpha \lambda^{-2} , \\
	\sigma_{11}^\text{v}/\eta_\alpha &= \lambda^{-2}\, \mathfrak{D}^\alpha \lambda^2 - \lambda\, \mathfrak{D}^\alpha \lambda^{-1} , \\
	\sigma_{11}^\text{v}/\eta_\alpha &= \lambda^{2}\, \mathfrak{D}^\alpha \lambda^2 - \lambda^{-1}\, \mathfrak{D}^\alpha \lambda^{-1} ,
	\end{aligned}
	\label{Tension}
\end{equation}
respectively.

In the limit of small tensile strains $\epsilon = 3(\lambda - 1)/K \simeq  0$, we obtain the linearised expression \eqref{Shear}\textsubscript{a} of the dimensionless tensile stress $\Sigma = \sigma_{11}/(\mu K)$ for all models. Therefore, the creep and relaxation behaviour in shear and tension-compression are formally equivalent at small strain amplitudes, see illustrations in Figure~\ref{fig:Creep}. Furthermore, we observe that $\sigma_{11}^\text{v}/\eta^\alpha$ has identical limits for Models~A and B as $\alpha \to 0$ or $\alpha \to 1$ (namely $0$ and $3\dot\lambda/\lambda$), but that Model~C has different limits (namely $\lambda^4-\lambda^{-2}$ and $(2\lambda^3+\lambda^{-3})\dot\lambda$). The above variability of the tensile viscous stress for large stretches provides a potential means of selecting practically relevant constitutive theories based on tensile stress relaxation experiments.

\begin{remark}
Upon division of $\sigma_{11}^\text{v}$ in Eq.~\eqref{Tension}\textsubscript{a} by the stretch $\lambda$, we recover the expression of the \textsubscript{`11'}-component of the first Piola-Kirchhoff stress $\bm{P}^\text{v} = \bm{\sigma}^\text{v} \bm{F}^{-\text{T}}$ provided in Eq.~(52) of Zhao et al. \cite{zhao21}{\,---\,}see also Eq.~(26) of Gao et al. \cite{gao23}. This equivalence is caused by the symmetry of the relative deformation gradient tensor $\bm{F}_{t|s}$ in the uniaxial tensile case, see definition in Eq.~\eqref{FRel}. Thus, this remark applies also to the pure shear and equibiaxial tensile motions for which the deformation gradient tensor can be chosen diagonal. We conclude that the experimental results obtained in \cite{zhao21, gao23} are consistent with Model~A, which is a frame-indifferent version of the theory proposed therein.
\end{remark}

\section{Incremental stress and acousto-elasticity}\label{sec:SoL}

\begin{figure}
	\centering
	\includegraphics{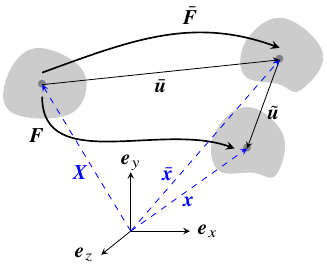}
	\caption{Acousto-elasticity. Combination of a large static deformation and a small incremental perturbation. \label{fig:SoL}}
\end{figure}

In this section, the material is subjected to an infinitesimal perturbation $\tilde{\bm u} = \bm{x}-\bar{\bm x}$ of a motionless equilibrium $\bar{\bm x}$ whose stress $\bar{\bm\sigma} = -\bar{p}\bm{I} + \bar{\bm \sigma}^\text{e}$ is assumed homogeneous. Hence, the equilibrium equation $\bar\nabla \cdot \bar{\bm\sigma} = \bm{0}$ for the pre-deformation $\bar{\bm x} = \bar{\bm F} {\bm X}$ is naturally satisfied. Here, quantities with an overbar are related to the statically pre-deformed configuration, whereas tildes mark infinitesimal increments, cf. Figure~\ref{fig:SoL}.

The total deformation gradient reads $\bm{F} = \bar{\bm F}+\bm{H}\bar{\bm F}$ where $\bm{H} = {\partial{\tilde{\bm u}}}/{\partial\bar{\bm x}}$ is the incremental displacement gradient, and the total particle velocity reduces to the incremental part: $\dot{\bm x} = \dot{\tilde{\bm{u}}} = \tilde{\bm v}$. 
Based on the decomposition $\bm{\sigma} = \bar{\bm\sigma} + \tilde{\bm\sigma}$ of the stress, linearisation of Cauchy's equation of motion with respect to $\tilde{\bm u}$ yields the incremental equations of motion  \cite{berjamin22},
\begin{equation}
	\rho\, \partial_t\tilde{\bm v} = \nabla\cdot \tilde{\bm\sigma},
	\label{incremental-motion}
\end{equation}
along with the linearised incompressibility constraint $\nabla\cdot \tilde{\bm v} = 0$.

By linearisation of \eqref{Constitutive}\textsubscript{b}, we arrive at the expression of the incremental stress $\tilde{\bm\sigma} = -\tilde{p}\bm{I} + \tilde{\bm\sigma}^\text{e} + \tilde{\bm\sigma}^\text{v}$, whose elastic part satisfies
\begin{equation}
\tilde{\bm\sigma}^\text{e} = 2\mathfrak{C}_+ (\bm{H}\bar{\bm B} + \bar{\bm B} \bm{H}^\text{T}) + 2\mathfrak{C}_- (\bm{H}^\text{T}\bar{\bm B}^{-1} + \bar{\bm B}^{-1}\bm{H}) , 
\end{equation}
see Eq.~\eqref{Mooney}.
Based on the relationship $\tilde{\bm \sigma}^\text{v} = \bar{\bm F}\tilde{\bm S}^\text{v}\bar{\bm F}^\text{T}$, Models~A-B and Model~C produce
\begin{equation}
	\tilde{\bm \sigma}^\text{v} = 2\mu\, \kappa * \tilde{\bm D} 
	,\qquad
	\tilde{\bm \sigma}^\text{v} = 2\mu\, \bar{\bm B}(\kappa * \tilde{\bm D})\bar{\bm B} ,
	\label{Increment}
\end{equation}
where $\tilde{\bm D} = \text{sym}\,\nabla \tilde{\bm v}$ is the symmetric part of the incremental velocity gradient. Note that Models~A and B yield the same viscous stress increment, \eqref{Increment}\textsubscript{a}.

\begin{remark}
We note that the non-objective fractional velocity gradient used by Delory et al. \cite{delory23} produces the same incremental stress as in Eq.~\eqref{Increment}\textsubscript{a}. Therefore, the Models~A and B can be used as a frame-indifferent substitute for the velocity gradient in the discussed study.
\end{remark}

Now consider  body waves propagating in a material subject to a homogeneous tri-axial stretch  $\bar{\bm F} = \text{diag}[\bar{\lambda}_1, \bar{\lambda}_2,\bar{\lambda}_3]$, with $\bar\lambda_1\bar\lambda_2\bar\lambda_3=1$ by incompressibility. The principal Cauchy stresses $\bar{\sigma}_i$ required to effect the pre-deformation are such that
\begin{equation}
\bar{\sigma}_i - \bar \sigma_j = 2\mathfrak{C}_+ (\bar\lambda_i^2 - \bar\lambda_j^2) - 2\mathfrak{C}_- (\bar\lambda_i^{-2} - \bar\lambda_j^{-2}) ,
\end{equation}
where the coefficients $\mathfrak{C}_\pm$ are defined in Eq.~\eqref{Mooney}.

We study harmonic principal body waves of the form $\tilde{\bm u} = \hat{\bm u}\, \text{e}^{\text{i}(\omega t - k x)}$, where $\hat{\bm u}$ is the constant amplitude  vector, $\omega$ is the angular frequency, $k$ is the wavenumber and $x$ is the direction of propagation.
We see from $\nabla\cdot \tilde{\bm v} = 0$ that the polarisation of the wave must be transverse to accommodate incremental incompressibility: $\hat{\bm u}\cdot \bm e_x = 0$.

Consider the wave polarised along $y$:  $\hat{\bm u} = \hat{u}\bm{e}_y$.
The harmonic amplitudes of $\partial_t\tilde{\bm v}$ and $\kappa*\tilde{\bm D}$ are $-\omega^2\hat{u} \bm{e}_y$ and $-(\text{i}\omega\tau)^\alpha \text{i}k\hat{u}\,\text{sym}(\bm{e}_y\otimes\bm{e}_x)$, respectively. 
Using the incremental equation of motion \eqref{incremental-motion}, we obtain the following dimensionless dynamic moduli $M^* = {\rho\omega^2}/({\mu k^2})$ for Models~A-B and for Model~C,
\begin{equation}
	\begin{aligned}
	M^* &= \tfrac{1+\beta}{2}\bar{\lambda}_1^2 + \tfrac{1-\beta}{2}\bar{\lambda}_2^{-2} + (\text{i}\omega\tau)^\alpha , \\
	M^* &= \tfrac{1+\beta}{2}\bar{\lambda}_1^2 + \tfrac{1-\beta}{2}\bar{\lambda}_2^{-2} + \bar{\lambda}_1^2\bar\lambda_2^2\, (\text{i}\omega\tau)^\alpha,
	\end{aligned}
	\label{Disp}
\end{equation}
respectively. The real and imaginary parts of $M^*$ represent the storage modulus $\mathfrak{Re}\, M^*$ and the loss modulus $\mathfrak{Im}\, M^*$, respectively.
We note that Models A and B cannot capture a change of loss modulus with the pre-stretch.

We can deduce other dispersion properties  from \eqref{Disp}, see Section~2.3 of Carcione \cite{carcione15}. For instance, the dissipation factors $d^* = \mathfrak{Im}\, M^* / \mathfrak{Re}\, M^*$ for Models A-B and for Model C are given by
\begin{equation}
	\begin{aligned}
	d^* &= \frac{(\omega \tau)^\alpha \sin(\alpha \tfrac\pi{2})}{\frac{1+\beta}{2}\bar{\lambda}_1^2 + \frac{1-\beta}{2}\bar{\lambda}_2^{-2} + (\omega \tau)^\alpha \cos(\alpha \tfrac\pi{2})}, \\[0.3pt]
	d^* &= \frac{\bar{\lambda}_1^2\bar\lambda_2^2\, (\omega \tau)^\alpha \sin(\alpha \tfrac\pi{2})}{\frac{1+\beta}{2}\bar{\lambda}_1^2 + \frac{1-\beta}{2}\bar{\lambda}_2^{-2} + \bar{\lambda}_1^2\bar\lambda_2^2\, (\omega \tau)^\alpha \cos(\alpha \tfrac\pi{2})} ,
	\end{aligned}
	\label{DissFact}
\end{equation}
respectively.
In the special case of neo-Hookean behaviour $\beta=1$, we see that the dissipation factor for Models~A-B is sensitive to the stretch along the propagation direction, while for Model~C, it depends on the stretch along the polarisation direction.
Upon nondimensionalisation,\footnote{We have $c = \mathfrak{Re}(1/\sqrt{M^*})^{-1}$ and $a = -\omega \tau \mathfrak{Im}(1/\sqrt{M^*})$.} the phase velocity $c = \omega / \mathfrak{Re}\, k$ and the attenuation factor $a = -\mathfrak{Im}\, k$ satisfy
\begin{equation}
	c^2 = \frac{2 \left(1+d^{*2}\right)}{1 + \sqrt{1+d^{*2}}}\, \mathfrak{Re}\, M^*, \qquad
	a^2 = \frac{(\omega \tau)^2}{\mathfrak{Re}\, M^*} \frac{\sqrt{1+d^{*2}} - 1}{2 \left(1+d^{*2}\right)} ,
	\label{Dispers}
\end{equation}
for all models.

For illustrative purposes, consider a neo-Hookean viscous body for which $\beta=1$, which is subjected to uni-axial traction of stretch $\bar \lambda$, see for instance the experimental configuration in Delory et al. \cite{delory24}.
Three types of principal waves may propagate in such a deformed body \cite{destrade10c}. 
First, the wave travelling in a direction perpendicular to the uni-axial tension, and polarised along that direction (e.g., a wave propagating along the $x$-axis and polarised along the $y$-axis, which is the direction of uni-axial tension). Here, $\bar \lambda_1=\bar \lambda_3=\bar\lambda^{-1/2}$, $\bar\lambda_2=\bar\lambda$. Wave dispersion properties are then deduced from Eqs.~\eqref{Disp}-\eqref{Dispers}.

\begin{figure}[h]
	\begin{minipage}{0.49\textwidth}
		\centering
		(a)
		
		\includegraphics{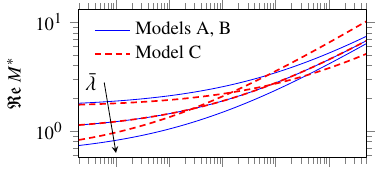}
		
		\vspace{-0.8em}
		
		\includegraphics{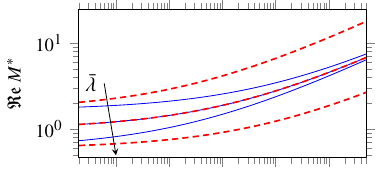}
		
		\vspace{-0.8em}
		
		\includegraphics{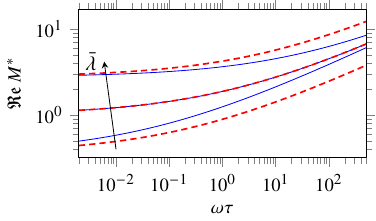}
	\end{minipage}\hfill
	\begin{minipage}{0.49\textwidth}
		\centering
		(b)
		
		\includegraphics{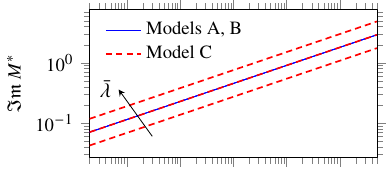}
		
		\vspace{-0.8em}
		
		\includegraphics{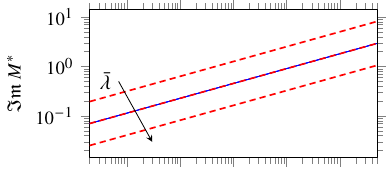}
		
		\vspace{-0.8em}
		
		\includegraphics{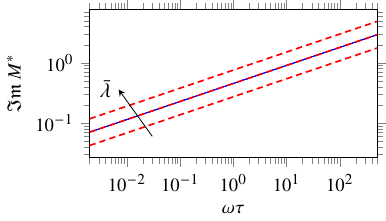}
	\end{minipage}
	
	\caption{Principal plane shear waves propagating in a neo-Hookean viscous material ($\beta=1$) subject to uni-axial tension. The material has fractional viscosity of order $\alpha = 0.3$ and the pre-stretch equals $\bar\lambda = \frac35, 1, \frac53$. Dimensionless storage modulus (a) and loss modulus (b) for waves polarised along the stretching direction (top), and for waves with propagation direction and polarisation transverse to the uni-axial tension (middle). Same for shear waves propagating along the stretching axis (bottom). \label{fig:SoLDisp}}
\end{figure}

In the first row of Figure~\ref{fig:SoLDisp} we show the variations of the corresponding  storage moduli $\mathfrak{Re}\, M^*$ and loss moduli $\mathfrak{Im}\, M^*$ with the dimensionless frequency $\omega \tau$ for some given values of pre-stretch, respectively found from
\begin{equation}
	M^* = \bar{\lambda}^{-1} + (\text{i}\omega\tau)^\alpha , \qquad
	M^* = \bar{\lambda}^{-1} + \bar{\lambda}\, (\text{i}\omega\tau)^\alpha .
	\label{Disp1}
\end{equation}
In particular, the figure shows the evolution of the loss modulus in terms of the frequency (Figure~\ref{fig:SoLDisp}b). As predicted earlier, the loss modulus is unaffected by variations of the pre-stretch for Models A and B, whereas the loss modulus increases with increasing values of $\bar\lambda$ when Model C is used.

\begin{figure}
	\begin{minipage}{0.32\textwidth}
		\centering
		(a)
		
		\includegraphics{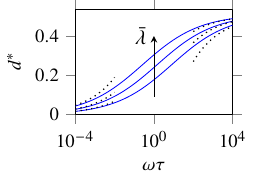}
	\end{minipage}\hfill
	\begin{minipage}{0.32\textwidth}
		\centering
		(b)
		
		\includegraphics{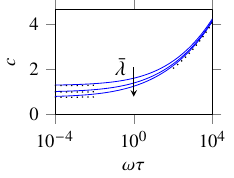}
	\end{minipage}\hfill
	\begin{minipage}{0.32\textwidth}
		\centering
		(c)
		
		\includegraphics{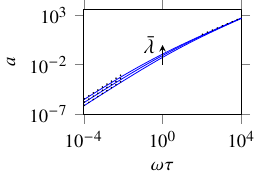}
	\end{minipage}
	
	\caption{Dispersion characteristics \eqref{DissFact}-\eqref{DissLim} in terms of the dimensionless frequency: (a) dissipation factor, (b) normalised phase velocity, (c) normalised attenuation coefficient. Models~A and B with $\alpha = 0.3$, $\beta=1$ and stretch $\bar\lambda = \frac35, 1, \frac53$, where shear waves are polarised along the stretching direction. \label{fig:DispProp}}
\end{figure}

For Models A and B{\,---\,}that is, for $M^*$ satisfying \eqref{Disp1}\textsubscript{a}{\,---\,}the relationships \eqref{DissFact}-\eqref{Dispers} entail the following asymptotic expansions for $0<\alpha<1$,
\begin{equation}
	{\renewcommand{\arraystretch}{1.4}
	\begin{array}{ll}
		d^* \underset{0}{\sim} (\omega \tau)^\alpha \sin(\alpha \tfrac\pi{2})\, \bar{\lambda}, \quad &
		d^* \underset{\infty}{\sim} \left(1 - (\omega \tau)^{-\alpha} \sec(\alpha \tfrac\pi{2}) / \bar{\lambda}\right) \tan(\alpha \tfrac\pi{2}) ,\\
		c \underset{0}{\sim} \left(1 + (\omega \tau)^{2\alpha} \sin(\alpha \tfrac\pi{2})^2\right)  \bar{\lambda}^{-\frac12} , \quad &
		c \underset{\infty}{\sim} (\omega \tau)^{\frac\alpha2} \sec(\alpha \tfrac\pi{4}) ,
		\\
		a \underset{0}{\sim} \tfrac12 (\omega\tau)^{1+\alpha} \sin(\alpha \tfrac\pi{2})\, \bar{\lambda}^{\frac32}, \quad &
		a \underset{\infty}{\sim} (\omega\tau)^{1-\frac\alpha2}\sin(\alpha\tfrac{\pi}{4}) ,
	\end{array}}
	\label{DissLim}
\end{equation} 
at low and high dimensionless frequency $\omega \tau$.
The above quantities \eqref{DissFact}\textsubscript{a}-\eqref{Dispers} with $\beta=1$ and the approximations \eqref{DissLim} are displayed in Figure~\ref{fig:DispProp} for several values of the pre-stretch. 
Although  the loss modulus does not vary with pre-stretch, $\bar\lambda$ still influences the dispersion properties through modification of the storage modulus. Of course, similar computations can be carried out for viscous Mooney-Rivlin solids $-1\leq \beta < 1$, for Model~C, and even with other wave polarisations, given that Eq.~\eqref{Dispers} is model-independent.

The second principal shear wave propagates and is polarised transversely to the direction of stretching.
Then $\bar \lambda_1 = \bar\lambda_2=\bar \lambda^{-1/2}$, $\bar\lambda_3=\bar \lambda$, and
\begin{equation}
	M^* = \bar{\lambda}^{-1} + (\text{i}\omega\tau)^\alpha , \qquad
	M^* = \bar{\lambda}^{-1} + \bar{\lambda}^{-2} (\text{i}\omega\tau)^\alpha ,
	\label{Disp2}
\end{equation}
for Models~A-B and for Model C, respectively.
Finally, the third wave propagates along the direction of stretching and $\bar \lambda_1 =\bar \lambda$, $ \bar\lambda_2=\bar\lambda_3=\bar \lambda^{-1/2}$.
In this case, we have the following dimensionless dynamic moduli for Models~A-B and for Model C,
\begin{equation}
	M^* = \bar{\lambda}^{2} + (\text{i}\omega\tau)^\alpha , \qquad
	M^* = \bar{\lambda}^{2} + \bar{\lambda}\, (\text{i}\omega\tau)^\alpha,
	\label{Disp3}
\end{equation}
respectively,
see the second and third rows of Figure~\ref{fig:SoLDisp} for the variations of the storage and loss moduli with the frequency when $\alpha=0.3$.

As shown in the figure, the first two waves yield a decrease of the storage modulus with increasing values of the applied stretch (see the first two rows of Figure~\ref{fig:SoLDisp}), whereas the third wave leads to the opposite trend (third row of Figure~\ref{fig:SoLDisp}). This feature can be explained by the relationship between the propagation direction and the stretching direction. In fact, the first two waves both propagate orthogonally to the direction of stretching. Hence, by virtue of incompressibility, they are subject to a contraction along their propagation direction if $\bar\lambda >1$. In contrast, the third wave travels along the direction of stretching, which implies that this wave is subject to an extension along its propagation direction if $\bar\lambda >1$.


\section{Conclusion}\label{sec:Conclu}

Several fractional viscous stresses were investigated. Among others, Models A, B and C are both physically satisfactory from the points of view of material frame-indifference and thermodynamic consistency. 
Nevertheless, one might prefer \emph{Model~A} or \emph{Model~B} due to their ease of use (e.g., for the study of shearing and tensile motions in creep and relaxation), and for the mathematical properties of the viscous limit. Although Models A and B both correspond to the classical Newtonian theory of viscosity for $\alpha \to 1$, their elastic limits $\alpha \to 0$ differ. In passing, we observe that linear combinations of these models produce Mooney-Rivlin stresses \eqref{Mooney} in the elastic limit, see the expressions in Table~\ref{tab:Constit}.

Despite this observation, it appears that Models A and B are very similar in many respects. In fact, they lead to the same creep and relaxation behaviour in simple shear. They produce also the same incremental stresses, as shown by the theory of acoustoelasticity. Nevertheless, they can be distinguished according to their creep and relaxation behaviour at large stretches \eqref{Tension}, which could be useful for experimental calibration purposes.

Interestingly, we note that the dispersion relationships obtained for body waves in Section~\ref{sec:SoL} are reminiscent of experimental and theoretical results obtained for Lamb wave propagation in stretched plates \cite{delory23}, as well as for wave propagation in stretched strips \cite{delory24}. On this basis, an experimentally calibrated theory that complies with the elementary principles of physics has yet to be established. The present study provides relevant models to carry out such a task.

On the one hand, if the loss modulus does not vary significantly with applied stretches, then Models A or B might be satisfactory. On the other hand, if the experimental data shows that the loss modulus varies with the pre-stretch, then Model~C might be preferred, despite its mathematical complexity for elementary shearing and tensile motions.
Alternatively, a more general fractional viscosity theory based on Model~A or Model~B could be derived by including further tensor invariants \cite{destrade09} (e.g., in the framework of the K-BKZ theory), especially if it provides better fit with experiments. Such an approach was followed by Delory et al. \cite{delory23} for the modelling of their experimental data.

Some limitations of the present study are the restriction to incompressible motions, and to fractional viscosity theories with one differential order only. Related compressible theories could be obtained by removing the incompressibility constraint, and by incorporating the missing strain and strain-rate tensor invariants. In the same spirit, fractional Maxwell models with two fractional differential orders could be developed \cite{palade99}.

\subsection*{Dedication}

This paper is dedicated to Ray Ogden, a mentor and an exemplar model.

\subsection*{Declaration of conflicting interests}
	The Authors declare that there is no conflict of interest.

\subsection*{Funding}
	This project has received funding from the European Union's Horizon 2020 research and innovation programme under grant agreement TBI-WAVES --- H2020-MSCA-IF-2020 project No. 101023950.

\addcontentsline{toc}{section}{References}

\bibliography{biblio}{}


\end{document}